\documentclass[prd,twocolumn,tightenlines]{revtex4}
\usepackage{amsmath}
\usepackage{amssymb}
\usepackage{graphicx}

\newcommand{\be}{\begin{eqnarray}}
\newcommand{\ee}{\end{eqnarray}}

 \newcommand{\gsim}{\mathrel{\hbox{\rlap{\lower.55ex \hbox {$\sim$}}
                   \kern-.3em \raise.4ex \hbox{$>$}}}}
\newcommand{\lsim}{\mathrel{\hbox{\rlap{\lower.55ex \hbox {$\sim$}}
                   \kern-.3em \raise.4ex \hbox{$<$}}}}

\begin{document}

\title{Production of large transverse momentum dileptons and photons in $pp$, $dA$ and $AA$ collisions by photoproduction processes }
\author{ Yong-Ping Fu$^{1)}$ and Yun-De Li$^{2)}$ \\ Department of Physics, Yunnan University, Kunming 650091, China\\
$1)$ ynufyp@sina.cn; $2)$ yndxlyd@163.com }
\date{\today}

\begin{abstract}
The production of large $P_{T}$ dileptons and photons originating
from photoproduction processes in $pp$, $dA$ and $AA$ collisions is
calculated. We find that the contribution of dileptons and photons
produced by photoproduction processes is not prominent at RHIC
energies. However, the numerical results indicate that the
modification of photoproduction processes becomes evident in the
large $P_{T}$ region for $pp$, $dA$ and $AA$ collisions at LHC
energies.


\end{abstract}

\maketitle

\address {Department of Physics, Yunnan University, Kunming 650091, People's
Republic of China }

\section{Introduction}
Hadronic processes for producing large transverse momentum($P_{T}$)
dileptons and photons are very important in the research of
relativistic $pp$, $dA$ and $AA$ collisions. Since photons and
dileptons do not participate in the strong interaction directly, the
photon or dilepton production can test the predictions of pQCD
calculations, and probe the strong interacting matter(quark-gluon
plasma, QGP). The hard scattering of partons is a well-known source
of large $P_{T}$ dileptons and photons in relativistic hadronic
collisions. The photons(and dileptons) are produced from various
processes in relativistic $AA$ collisions(relativistic heavy ion
collisions): primary hard photons from initial parton collisions
\cite{1,2,12,3,4,4a,4b,5,6,7,7.1}, thermal photons from the QGP
\cite{8,9,9a,10,13,15,16,17,17a,17b} and hadronic gas
\cite{11,11a,11b,11c,11d}, photons from the jet-photon conversion in
the thermal medium \cite{18,19,19.1,19.2}, and photons from hadronic
decays after freeze-out \cite{11e}. In relativistic $AA$ collisions
the contribution of photons produced by the jet-photon conversion in
the thermal medium is also important in the large $P_{T}$ region
\cite{18,19}.

\begin{figure*}
\includegraphics[scale=0.6]{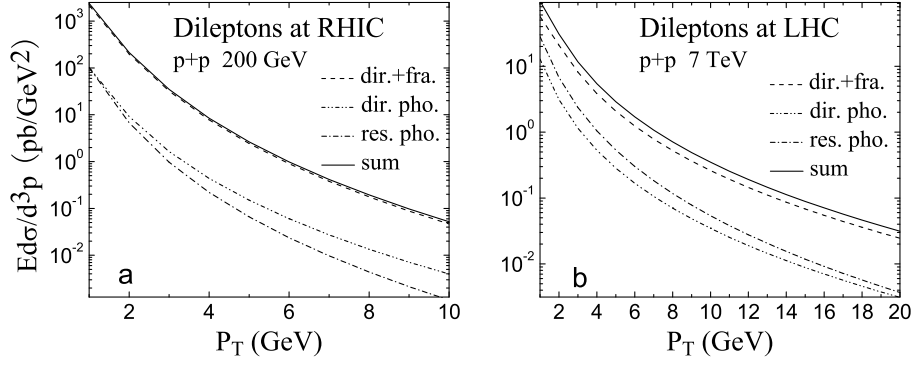}
\caption{\label{fig1} a: Invariant cross section of dileptons for
y=0 in $p+p$ collisions at $\sqrt{s}$=200 GeV. (Dash line)The sum of
direct dileptons(dir.) and fragmentation dileptons(fra.). (Dash dot
dot line)Dileptons produced by direct photoproduction
processes(dir.pho.). (Dash dot line)Dileptons produced by resolved
photoproduction processes(res.pho.). (Solid line)The sum of direct
dileptons, fragmentation dileptons and dileptons produced by
photoproduction processes. b: Same as panel a but for $p+p$
collisions at $\sqrt{s}$=7 TeV.}
\end{figure*}

In the present work, we investigate the production of large $P_{T}$
dileptons and photons induced by photoproduction processes in $pp$,
$dA$ and $AA$ collisions at Relativistic Heavy Ion Collider(RHIC)
and Large Hadron Collider(LHC) energies. The photoproduction
processes play a fundamental role in the $ep$ deep inelastic
scattering at Hadron Electron Ring Accelerator(HERA)
\cite{21,22,23,23a}. In photoproduction processes of the $ep$ deep
inelastic scattering, a high energy photon emitted from the incident
electron directly interacts with the proton by the interaction of
$\gamma p\rightarrow jets$. Besides, the uncertainty principle
allows the high energy photon for a short time to fluctuate into a
quark-antiquark pair which then interacts with the partons of the
proton. In such interactions the resolved photon can be regarded as
an extended object consisting of quarks and also gluons. The
interactions are the so-called resolved photoproduction processes.

We extend the photoproduction mechanism to the photon and dilepton
production in $pp$, $dA$ and $AA$ collisions. Charged partons of the
incident nucleon also can emit high energy photons(and resolved
photons) in relativistic hadron-hadron, hadron-nucleus and
nucleus-nucleus collisions. The photon spectrum from the charged
parton is given by \cite{24,25}
\begin{eqnarray}\label{pho}
f_{\gamma/q}(z)=e^{2}_{q}\frac{\alpha}{2\pi}
\frac{1+(1-z)^{2}}{z}\ln\left(\frac{Q^{2}_{1}}{Q_{2}^{2}}\right),
\end{eqnarray}
where $\alpha$ is the electromagnetic coupling parameter, $z$ is the
momentum ratio of the photon energy and the energy of the quark, the
values $Q_{1}^{2}$ and $Q_{2}^{2}$ stand for the maximum and minimum
value of the momentum transfer, respectively. In direct
photoproduction processes, the high energy photon emitted from the
charged parton of the incident nucleon interacts with the parton of
another incident nucleon by the interaction of $q\gamma\rightarrow
q\gamma^{*}$(or $\gamma$). In resolved photoproduction processes,
the hadron-like photon interacts with the parton of the nucleon by
the interactions of $q_{\gamma}\bar{q}\rightarrow g\gamma^{*}$,
$q_{\gamma}g\rightarrow q\gamma^{*}$ and $g_{\gamma}q\rightarrow
q\gamma^{*}$, here $q_{\gamma}$($g_{\gamma}$) denotes the parton of
the resolved photon.

The paper is organized as follows. In Sec.II we present the
production of large $P_{T}$ dileptons and photons in hadronic
collisions. The direct and resolved photoproduction processes are
presented. In Sec.III we briefly review the production of thermal
dileptons and photons in the QGP. In Sec.IV the production rate of
jet-dilepton(photon) conversion is discussed. The numerical results
at RHIC and LHC energies are plotted in Sec.V. Finally, the summary
is given in Sec.VI.

\section{Large $P_{T}$ dilepton and photon production}
\subsection{Large $P_{T}$ dilepton production}

The large $P_{T}$ dileptons produced by initial parton collisions
can be divided into two categories: direct dileptons produced by the
annihilation and Compton scattering of partons, fragmentation
dileptons produced by the bremsstrahlung emitted from final state
partons \cite{3,4,4a}. The direct dileptons ($dir.l^{+}l^{-}$)
produced by the subprocesses $q\bar{q}\rightarrow
g(\gamma^{*}\rightarrow l^{+}l^{-})$ and $qg\rightarrow q
(\gamma^{*}\rightarrow l^{+}l^{-})$ in the hadronic
collisions($AB\rightarrow l^{-}l^{+}X$) satisfy the following
invariant cross section \cite{2,3,4,4a}
\begin{eqnarray}
\frac{d\sigma_{dir.l^{+}l^{-}}}{dM^{2}dP^{2}_{T}dy}\label{eq005}&=&\frac{1}{\pi}\int
dx_{a}G_{a/A}(x_{a},Q^{2})G_{b/B}(x_{b},Q^{2})
\nonumber\\[1mm]
&&\times
\frac{x_{a}x_{b}}{x_{a}-x_{1}}\frac{d\hat{\sigma}}{dM^{2}d\hat{t}}(x_{a},x_{b},P_{T},M^{2}),
\end{eqnarray}
where the functions $G_{a/A}(x_{a},Q^{2})$ and
$G_{b/B}(x_{b},Q^{2})$ are parton distributions of nucleons, $x_{a}$
and $x_{b}$ are the parton's momentum fraction. We have
$x_{b}=(x_{a}x_{2}-\tau)/(x_{a}-x_{1})$. The variables are
$x_{1}=\left(x_{T}^{2}+4\tau\right)^{1/2}e^{y}/2$, $
x_{2}=\left(x_{T}^{2}+4\tau\right)^{1/2}e^{-y}/2$,
$x_{T}=2P_{T}/\sqrt{s_{NN}}$ and $\tau=M^{2}/s_{NN}$. $y$ is the
rapidity, $M$ is the invariant mass of the lepton pair and
$\sqrt{s_{NN}}$ is the energy of the nucleon in the center-of-mass
system.

The cross section of the subprocesses $ab\rightarrow l^{+}l^{-}d$
with the invariant mass squared $M^{2}$ and Mandelstam variable
$\hat{t}$ can be written as \cite{2,3}
\begin{eqnarray}\label{sub001}
\frac{d\hat{\sigma}}{dM^{2}d\hat{t}}(ab\rightarrow l^{+}l^{-}
d)&=&\frac{\alpha}{3\pi
M^{2}}\sqrt{1-\frac{4m^{2}_{l}}{M^{2}}}\left(1+\frac{2m^{2}_{l}}{M^{2}}\right)
\nonumber\\[1mm]
&&\times \frac{d\hat{\sigma}}{d\hat{t}}(ab\rightarrow \gamma^{*} d),
\end{eqnarray}
where $m_{l}$ is the lepton mass. Here
$d\hat{\sigma}/d\hat{t}(ab\rightarrow \gamma^{*} d)$ denotes the
cross section of $q\bar{q}\rightarrow g\gamma^{*}$ and
$qg\rightarrow q \gamma^{*}$ \cite{2}.

The parton distribution $G_{i/A}(x_{i},Q^{2})$ of the nucleon is
given by \cite{5}
\begin{eqnarray}
G_{i/A}(x_{i},Q^{2})\!\!=\!\!R_{i/A}(x_{i},Q^{2})
\!\!\left[\!\frac{Z}{A}p_{i}(x_{i},Q^{2})\!\!+\!\!\frac{N}{A}n_{i}(x_{i},Q^{2})\right]\!\!,
\end{eqnarray}
where $R_{i/A}(x_{i},Q^{2})$ is the nuclear modification factor
\cite{7.1}, $Z$ is the proton number, $N$ is the neutron number and
$A$ is the nucleon number. $p_{i}(x_{i},Q^{2})$ and
$n_{i}(x_{i},Q^{2})$ are the parton distributions of protons and
neutrons, respectively. We choose the momentum scale as
$Q^{2}=4P_{T}^{2}$.

\begin{figure*}
\includegraphics[scale=0.6]{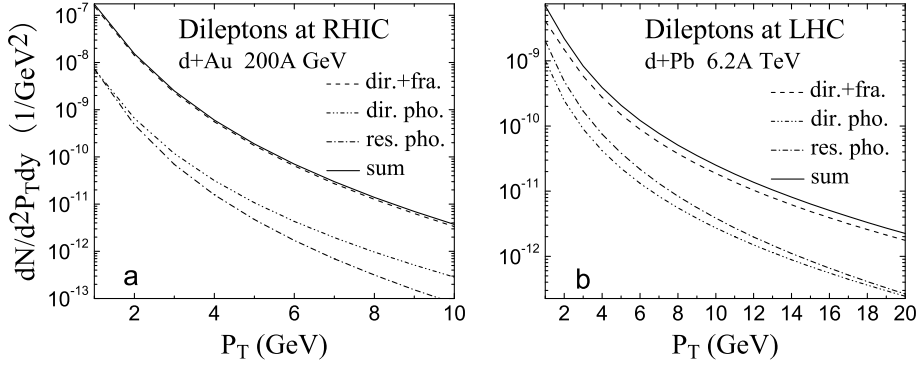}
\caption{\label{fig2} Dilepton yield for y=0 in central $d$+Au
collisions at $\sqrt{s_{NN}}=$200 GeV(panel a) and $d$+Pb collisions
at $\sqrt{s_{NN}}=$6.2 TeV(panel b).}
\end{figure*}

The fragmentation dileptons ($fra.l^{+}l^{-}$) are produced by the
hard scattering $ab\rightarrow (c\rightarrow xl^{+}l^{-})d$
\cite{2,3,4,4a}. The invariant cross section of fragmentation
dileptons is
\begin{eqnarray}\label{dirll}
\frac{d\sigma_{fra.l^{+}l^{-}}}{dM^{2}dP^{2}_{T}dy}\label{eq006}&\!\!=&\!\!\frac{1}{\pi}\int
dx_{a}\int dx_{b} G_{a/A}(x_{a},Q^{2}) G_{b/B}(x_{b},Q^{2})
\nonumber\\[1mm]
&&\times
D_{q_{c}}^{l^{+}l^{-}}(z_{c},Q^{2})\frac{x_{a}x_{b}}{z_{c}(x_{a}x_{b}-\tau)}\nonumber\\[1mm]
&&\times\frac{d\hat{\sigma}_{par.}}{d\hat{t}}(x_{a},x_{b},z_{c},P_{T},M^{2}),
\end{eqnarray} 
where $ z_{c}=(x_{a}x_{2}+x_{b}x_{1})/(x_{a}x_{b}-\tau)$ is the
momentum fraction of the final state dilepton. The dilepton
fragmentation function is given by
\begin{eqnarray}
D_{q_{c}}^{l^{+}l^{-}}(z_{c},M^{2},Q^{2})&=&\frac{\alpha}{3\pi
M^{2}}\sqrt{1-\frac{4m^{2}_{l}}{M^{2}}}\left(1+\frac{2m^{2}_{l}}{M^{2}}\right)
\nonumber\\[1mm]
&&\times D^{\gamma^{*}}_{q_{c}}(z_{c},Q^{2}),
\end{eqnarray}
where $D^{\gamma^{*}}_{q_{c}}(z_{c},Q^{2})$ is the virtual photon
fragmentation function \cite{3}. $d\hat{\sigma}_{par.}/d\hat{t}$
denotes the cross section of the subprocesses. These subprocesses
are $qq'\rightarrow qq'$, $q\bar{q}'\rightarrow q\bar{q}'$,
$qq\rightarrow qq$, $q\bar{q}\rightarrow q'\bar{q}'$,
$q\bar{q}\rightarrow q\bar{q}$, $gg\rightarrow q\bar{q}$,
$qg\rightarrow qg$, $q\bar{q}\rightarrow gg$ and $gg\rightarrow gg$
\cite{1}.

\subsection{Photoproduction processes in large $P_{T}$ dilepton production}

In direct photoproduction processes, the parton $a$ of the incident
nucleon $A$ can emit a large $P_{T}$ photon, then the high energy
photon interacts with the parton $b$ of another incident nucleon $B$
by the interaction of $q_{b}\gamma\rightarrow
q(\gamma^{*}\rightarrow l^{+}l^{-})$. The invariant cross section of
large $P_{T}$ dileptons produced by direct photoproduction processes
(dir. pho.) is given by
\begin{eqnarray}\label{dirpho}
\frac{d\sigma_{dir.pho.}}{dM^{2}dP^{2}_{T}dy}\label{eq008}\!\!&=&\!\!\frac{2}{\pi}\int
dx_{a}\int
dx_{b}G_{a/A}(x_{a},Q^{2})G_{b/B}(x_{b},Q^{2})\nonumber\\[1mm]
&&\times
f_{\gamma/q_{a}}(z_{a})\frac{x_{a}x_{b}z_{a}}{x_{a}x_{b}-x_{a}x_{2}}\nonumber\\[1mm]
&&\times
\frac{d\hat{\sigma}}{dM^{2}d\hat{t}}(x_{a},x_{b},z_{a},P_{T},M^{2}),
\end{eqnarray} 
where $f_{\gamma/q_{a}}(z_{a})$ is the photon spectrum from the
quark. According to \cite{25} we choose $Q_{1}^{2}$ to be the
maximum value of the momentum transfer given by
$\hat{s}/4-m_{l}^{2}$ and the choice of the momentum transfer
$Q_{2}^{2}=$1 GeV$^{2}$ is made such that the photon is sufficiently
off shell for the parton model to be applicable.
$\hat{s}=x_{a}x_{b}z_{a}s_{NN}$ is the square of the center-of-mass
energy for the subprocesses. The function
$d\hat{\sigma}/dM^{2}d\hat{t}$ denotes the cross section of
subprocess $q\gamma\rightarrow q(\gamma^{*}\rightarrow l^{+}l^{-})$
\cite{2}. Here $z_{a}=(x_{b}x_{1}-\tau)/(x_{a}x_{b}-x_{a}x_{2})$ is
the momentum fraction of the photon emitted from the quark of the
nucleon.

In resolved photoproduction processes, the parton $a$ of the
incident nucleon $A$ emits a high energy resolved photon, then the
parton $a'$ of the resolved photon interacts with the parton $b$ of
another incident nucleon $B$ by the interactions of
$q_{a'}\bar{q}_{b}\rightarrow g \gamma^{*}$, $q_{a'}g_{b}\rightarrow
q
 \gamma^{*}$ and $q_{b}g_{a'}\rightarrow
q
 \gamma^{*}$. The invariant cross section of large $P_{T}$
dileptons produced by resolved photoproduction processes (res. pho.)
can be written as
\begin{eqnarray}\label{respho}
\frac{d\sigma_{res.pho.}}{dM^{2}dP_{T}^{2}dy}\label{eq009}&=&\frac{2}{\pi}\int
dx_{a}\int dx_{b}\int dz_{a'}G_{a/A}(x_{a},Q^{2})
\nonumber\\[1mm]
&&\times
G_{b/B}(x_{b},Q^{2})f_{\gamma/q_{a}}(z_{a})G_{q_{a'}/\gamma}(z_{a'},Q^{2})\nonumber\\[1mm]
&&\times\frac{x_{a}x_{b}z_{a}z_{a'}}{x_{a}x_{b}z_{a'}-x_{a}z_{a'}x_{2}}
\nonumber\\[1mm]
&&\times
\frac{d\hat{\sigma}}{dM^{2}d\hat{t}}(x_{a},x_{b},z_{a},z_{a'},P_{T},M^{2}),
\end{eqnarray}
where $G_{q_{a'}/\gamma}(z_{a'},Q^{2})$ is the parton distribution
of the resolved photon \cite{7}. The cross section
$d\hat{\sigma}/dM^{2}d\hat{t}$ of $q\bar{q}\rightarrow
g(\gamma^{*}\rightarrow l^{+}l^{-})$ and $qg\rightarrow q
(\gamma^{*}\rightarrow l^{+}l^{-})$ is discussed in
Eq.(\ref{sub001}). The variable $z_{a'}$ denotes the momentum
fraction of the parton of the resolved photon emitted from the
quark. Here we have
$z_{a}=(x_{b}x_{1}-\tau)/(x_{a}x_{b}z_{a'}-x_{a}z_{a'}x_{2})$ and
$\hat{s}=x_{a}x_{b}z_{a}z_{a'}s_{NN}$.

\subsection{Real photon production}

Because a virtual photon can directly decay into a dilepton, the
invariant cross sections of large $P_{T}$ photons can be derived
from the cross sections of the dilepton production if the invariant
mass of the lepton pair is zero($M^{2}=0$). The maximum momentum
transfer $Q_{1}^{2}$ in Eq.(\ref{pho}) is $\hat{s}/4$ in the real
photon production. The prompt photons are produced by the direct
production ($dir.\gamma$) and the fragmentation process
($fra.\gamma$). The invariant cross section for direct photons is
given by \cite{1,2,12}
\begin{eqnarray}\label{dirr}
E\frac{d\sigma_{dir.\gamma}}{d^{3}P}\label{r1}&=&\frac{1}{\pi}\int
dx_{a}G_{a/A}(x_{a},Q^{2})G_{b/B}(x_{b},Q^{2})
\frac{x_{a}x_{b}}{x_{a}-x_{1}}\nonumber\\[1mm]
&&\times \frac{d\hat{\sigma}}{d\hat{t}}(x_{a},x_{b},P_{T}),
\end{eqnarray}
where $x_{b}=x_{a}x_{2}/(x_{a}-x_{1})$. In the real photon case we
have $x_{1}=x_{T}e^{y}/2$ and $ x_{2}=x_{T}e^{-y}/2$. The function
$d\hat{\sigma}/d\hat{t}$ of Eq.(\ref{dirr}) denotes the cross
section of subprocesses $q\bar{q}\rightarrow g\gamma$ and
$qg\rightarrow q \gamma$ \cite{1}. The invariant cross section for
fragmentation photons is given by \cite{1,2,12}
\begin{eqnarray}\label{fragr}
E\frac{d\sigma_{fra.\gamma}}{d^{3}P}\label{eq002}&=&\frac{1}{\pi}\int
dx_{a}\int
dx_{b}G_{a/A}(x_{a},Q^{2})G_{b/B}(x_{b},Q^{2})\nonumber\\[1mm]
&&\times \frac{D_{q_{c}}^{\gamma}(z_{c},Q^{2})}{
z_{c}}\frac{d\hat{\sigma}_{par.}}{d\hat{t}}(x_{a},x_{b},z_{c},P_{T}),
\end{eqnarray}
where $D_{q_{c}}^{\gamma}(z_{c},Q^{2})$ is the real photon
fragmentation function and $z_{c}=x_{1}/x_{a}+x_{2}/x_{b}$
\cite{1,12}. The cross section $d\hat{\sigma}_{par.}/d\hat{t}$ of
Eq.(\ref{fragr}) is discussed in Eq.(\ref{dirll}).

\begin{figure*}
\includegraphics[scale=0.6]{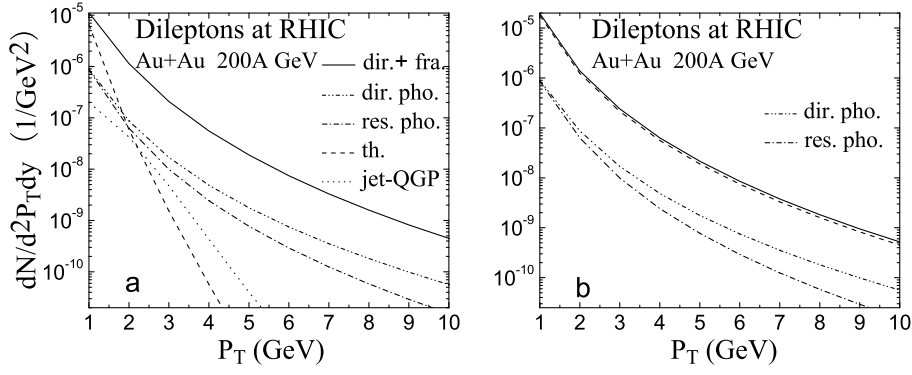}
\caption{\label{fig3} a: Dilepton yield for y=0 in central Au+Au
collisions at $\sqrt{s_{NN}}$=200 GeV. (Solid line)The sum of direct
dileptons(dir.) and fragmentation dileptons(fra.). (Dash dot dot
line)Dileptons produced by direct photoproduction
processes(dir.pho.). (Dash dot line)Dileptons produced by resolved
photoproduction processes(res.pho.). (Dash line)Thermal
dileptons(th.) produced by the QGP. (Dot line)The jet-dilepton
conversion(jet-QGP) in the plasma. b: The contribution of
photoproduction processes at RHIC energies. (Dash line)The sum of
direct dileptons, fragmentation dileptons, thermal dileptons
produced by the QGP and dileptons from the jet-dilepton conversion.
(Solid line)The sum of dash line, dash dot line and dash dot dot
line.}
\end{figure*}

The invariant cross section of real photons produced by direct
photoproduction processes is
\begin{eqnarray}\label{dirphor}
E\frac{d\sigma_{dir.pho.}}{d^{3}P}&=&\frac{2}{\pi}\int dx_{a}\int
dx_{b}G_{a/A}(x_{a},Q^{2})G_{b/B}(x_{b},Q^{2})\nonumber\\[1mm]
&&\times
f_{\gamma/q_{a}}(z_{a})\frac{x_{a}x_{b}z_{a}}{x_{a}x_{b}-x_{a}x_{2}}\nonumber\\[1mm]
&&\times
 \frac{d\hat{\sigma}}{d\hat{t}}(x_{a},x_{b},z_{a},P_{T}),
\end{eqnarray}
where $z_{a}=x_{b}x_{1}/(x_{a}x_{b}-x_{a}x_{2})$. The real photon
production of resolved photoproduction processes is
\begin{eqnarray}\label{resphor}
E\frac{d\sigma_{res.pho.}}{d^{3}P}&=&\frac{2}{\pi}\int dx_{a}\int
dx_{b}\int dz_{a'}G_{a/A}(x_{a},Q^{2})
\nonumber\\[1mm]
&&\times
G_{b/B}(x_{b},Q^{2})f_{\gamma/q_{a}}(z_{a})G_{q_{a'}/\gamma}(z_{a'},Q^{2})\nonumber\\[1mm]
&&\times\frac{x_{a}x_{b}z_{a}z_{a'}}{x_{a}x_{b}z_{a'}-x_{a}z_{a'}x_{2}}
\nonumber\\[1mm]
&&\times
\frac{d\hat{\sigma}}{d\hat{t}}(x_{a},x_{b},z_{a},z_{a'},P_{T}),
\end{eqnarray}
where the elementary cross sections $d\hat{\sigma}/d\hat{t}$ of
Eq.(\ref{dirphor}) and Eq.(\ref{resphor}) are similar to the cases
of the dilepton production in Eq.(\ref{dirpho}) and
(\ref{respho})(but with $M^{2}$=0), respectively. Here we have
$z_{a}=x_{b}x_{1}/(x_{a}x_{b}z_{a'}-x_{a}z_{a'}x_{2})$ for
Eq.(\ref{resphor}).

\section{Production of thermal dileptons and photons}

The yield of thermal dileptons ($th.l^{+}l^{-}$) with the low
dilepton mass and large transverse momentum can be written as
\cite{8,10,11}
\begin{eqnarray}
\frac{dN_{th.l^{+}l^{-}}}{dM^{2}dP^{2}_{T}dy}\label{eq011}&=&\pi
R_{A}^{2}\frac{\sigma_{q\bar{q}}(M)}{4(2\pi)^{4}}M^{2}
\sqrt{\!1-\frac{4m_{q}^{2}}{M^{2}}}\frac{3\tau_{0}^{2}
T_{0}^{6}}{P_{T}^{6}}\nonumber\\[1mm]
&&\times
\!\!\left[G\!\!\left(\!\frac{P_{T}}{T_{0}}\!\right)\!\!-G\!\!\left(\!\frac{P_{T}}{T_{c}}\!\right)
\right] ,
\end{eqnarray} 
where $R_{A}$ is the nuclear radius, $m_{q}$ is the quark mass.
$\tau_{0}$ and $T_{0}$ are the initial time and the initial
temperature of the system, respectively. We use $\tau_{0}=0.26$
fm/$c$ for RHIC, $\tau_{0}=0.09$ fm/$c$ for LHC($\sqrt{s_{NN}}=2.76$
TeV) and $\tau_{0}=0.088$ fm/$c$ for LHC($\sqrt{s_{NN}}=5.5$ TeV).
The initial temperature of the QGP is chosen as $T_{0}=370$ MeV for
Au+Au collisions at $\sqrt{s_{NN}}$=200 GeV, $T_{0}=710$ MeV for
Pb+Pb collisions at $\sqrt{s_{NN}}$=2.76 TeV, and $T_{0}=845$ MeV
for Pb+Pb collisions at $\sqrt{s_{NN}}$=5.5 TeV. $T_{c}$(=160 MeV)
is the critical temperature of the phase transition \cite{12}. Here
$\sigma_{q\bar{q}}=4\pi\alpha^{2}N_{c}N_{s}^{2}e_{q}^{2}/3M^{2}$ is
the cross section of the process $q\bar{q}\rightarrow
\gamma^{*}\rightarrow l^{+}l^{-}$, $N_{c}(=3)$ is the color number,
$N_{s}(=2)$ is the spin number. The function $G(z)$ is given by
$G(z)=z^{3}(8+z^{2})K_{3}(z)$.

\begin{figure*}
\includegraphics[scale=0.6]{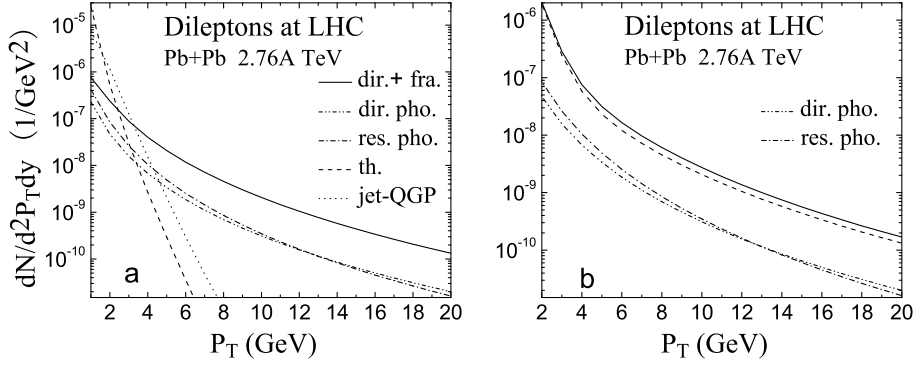}
\caption{\label{fig4} Same as Fig.\ref{fig3} but for central Pb+Pb
collisions at $\sqrt{s_{NN}}=$2.76 TeV.}
\end{figure*}

\begin{figure*}
\includegraphics[scale=0.6]{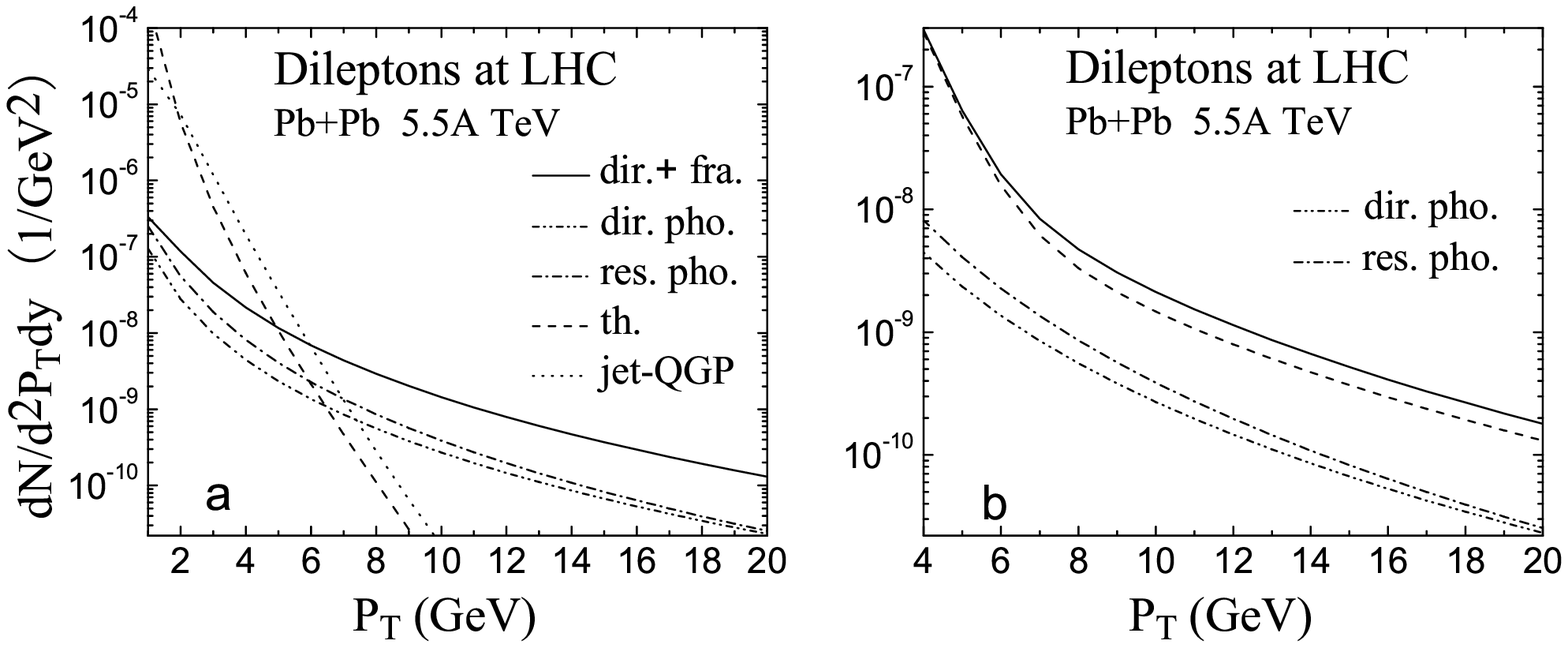}
\caption{\label{fig5}  Same as Fig.\ref{fig3} but for central Pb+Pb
collisions at $\sqrt{s_{NN}}=$5.5 TeV.}
\end{figure*}

The yield of thermal photons ($th.\gamma$) is given by the following
\cite{9,10}
\begin{eqnarray}
E\frac{dN_{th.\gamma}}{d^{3}P}\label{eq010}&=&\pi
R_{A}^{2}\frac{\alpha\alpha_{s}e_{q}^{2}}{4\pi^{2}}\int^{\tau
_{c}}_{\tau_{0}}\tau d\tau f_{th}(\emph{\textbf{p}}_{\gamma})T^{2}
\nonumber\\[1mm]
&&\times
\left[2\mathrm{ln}\left(\frac{6E_{\gamma}}{\pi\alpha_{s}T}\right)\!+C_{Com.}\!+C_{ann.}\right],
\end{eqnarray}
where $\tau_{c}=\tau_{0}(T_{0}/T_{c})^{3}$ is the critical time of
the phase transition. The parameters are $C_{Com.}=-0.416$ and
$C_{ann.}=-1.916$. $f_{th}$ is the thermal distribution of thermal
partons. In the Bjorken expansion, the temperature evolves as
$T=T_{0}(\tau_{0}/\tau)^{1/3}$ \cite{9a}.

\section{Jet-dilepton(photon) conversion}


The jet-dilepton conversion is induced by the annihilation of jets
passing through the QGP \cite{20a,20}. We rigorously derive the
dilepton production rate of the jet-dilepton conversion. Using the
relativistic kinetic theory the production rate can be written as
$R_{jet-l^{+}l^{-}}\!\!\!\!\propto\!\!1/(2\pi)^{6}\int\!\!
d^{3}\emph{\textbf{p}}_{1}\!\!\int\!\!
d^{3}\emph{\textbf{p}}_{2}f(\emph{\textbf{p}}_{1})f(\emph{\textbf{p}}_{2})v_{12}\sigma$,
where the relative velocity is $v_{12}=(p_{1}+ p_{2})^{2}/2p_{1}^{0}
p_{2}^{0}$. After some algebra the rate can be written as
\begin{eqnarray}
\frac{dR_{jet-l^{+}l^{-}}}{dM^{2}dP^{2}_{T}}\label{eqjet01}=\frac{\sigma_{q\bar{q}}
M^{2}}{2(2\pi)^{4}}\!\!\int dP'_{T}
\frac{f_{jet}(P'_{T})}{4P'_{T}}e^{-\frac{P_{T}^{2}}{4P'_{T}T}},
\end{eqnarray}
where $f_{jet}$ is the phase-space distribution of jets with the
large transverse momentum($P'_{T}$). If the jet distribution is
replaced by the thermal distribution $\exp(-E/T)=\exp(-P'_{T}\cosh
y/T)$, one can obtain the rate for producing thermal dileptons.

\begin{figure*}
\includegraphics[scale=0.6]{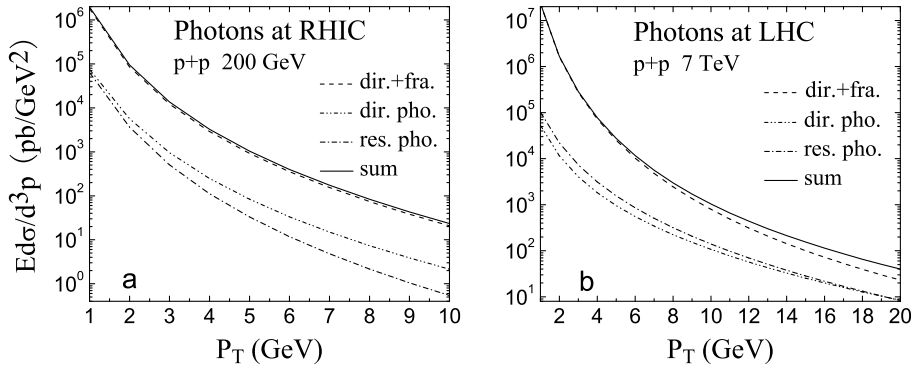}
\caption{\label{fig6}  Same as Fig.\ref{fig1} but for the real
photon production in $p+p$ collisions at RHIC(panel a) and LHC(panel
b) energies.}
\end{figure*}

\begin{figure*}
\includegraphics[scale=0.6]{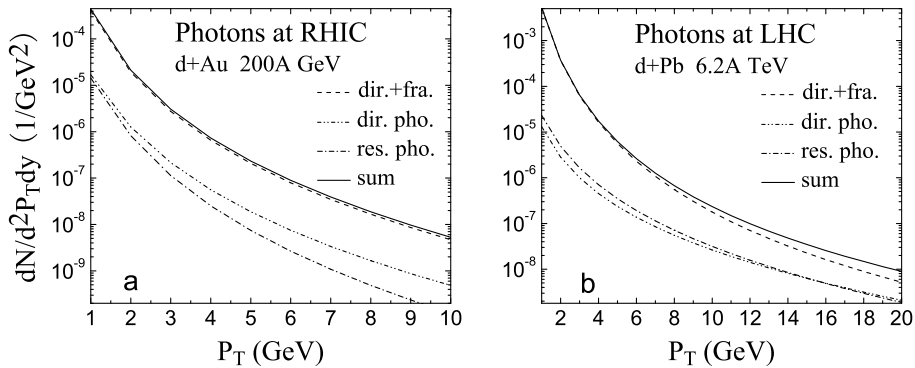}
\caption{\label{fig7} Same as Fig.\ref{fig2} but for the real photon
production in central $d$+Au collisions at RHIC(panel a) and $d$+Pb
collisions at LHC(panel b).}
\end{figure*}

The phase-space distribution of jets is given by \cite{18,19}
\begin{eqnarray}
f_{jet}(P'_{T})\label{eq015}&=&\frac{(2\pi)^{3}}{g\pi
R_{\bot}^{2}\tau
P'_{T}\cosh y}\frac{dN _{jet}}{d^{2}P'_{T}dy}\delta(\eta-y) \nonumber\\[1mm]
&&\times \Theta(\tau-\tau_{i})
\Theta(\tau_{max}-\tau)\Theta(R_{\bot}-r),
\end{eqnarray}
where $g(=6)$ is the spin and color degeneracy of quarks, $R_{\bot}$
is the transverse radius of the system, $\eta$ is the space-time
rapidity of the system, $\tau_{i}$ is the formation time for the
jet. We take $\tau_{max}$ as the smaller of the lifetime of the QGP
and the time taken by the jet produced at position $r$ to reach the
surface of the QGP. The yield of jets produced by $AA$ collisions
can be written as
\begin{eqnarray}
\frac{dN_{jet}}{d^{2}P'_{T}dy}=T_{AA}E'\frac{d\sigma_{jet}}{d^{3}P'}(y=0),
\end{eqnarray}
where the nuclear thickness $T_{AA}$ for zero impact parameter is
$9A^{2}/8\pi R^{2}_{\bot}$. The invariant cross section of the jet
production is given by
\begin{eqnarray}
E'\frac{d\sigma_{jet}}{d^{3}P'}&=&\frac{1}{\pi}\int
dx_{a}G_{a/A}(x_{a},Q^{2})G_{b/B}(x_{b},Q^{2})
\frac{x_{a}x_{b}}{x_{a}-x_{1}}\nonumber\\[1mm]
&&\times \frac{d\hat{\sigma}_{par.}}{d\hat{t}}(x_{a},x_{b},P'_{T}).
\end{eqnarray}

Jets passing through the QGP will lose energy. Induced gluon
bremsstrahlung, rather than elastic scattering of partons, is the
dominant mechanism of the jet energy loss \cite{12,18,19,EL1,EL2}.
Based on the AMY formulism, the energy loss of the final state
partons can be described as a dependence of the final state parton
spectrum $dN_{jet}/dE$ on time \cite{12,EL2.1}. Besides, the energy
loss of jets can be scaled as the square of the distance traveled
through the medium \cite{EL3}. Jets travel only a short distance
through the plasma before the jet-photon(or virtual photon)
conversion, and do not lose a significant amount of energy. The
energy loss effect of jets before they convert into photons(or
virtual photons) is found to be small, about 20$\%$ \cite{12,18,19}.

\begin{figure*}
\includegraphics[scale=0.6]{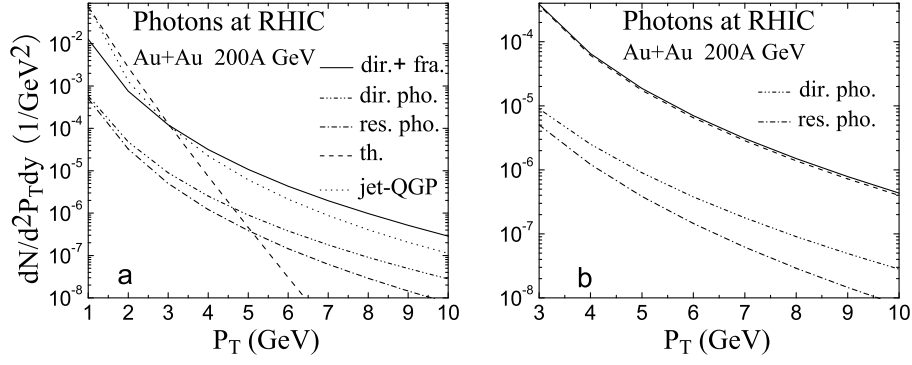}
\caption{\label{fig8}  Same as Fig.\ref{fig3} but for the real
photon production in central Au+Au collisions at RHIC energies.}
\end{figure*}

\begin{figure*}
\includegraphics[scale=0.6]{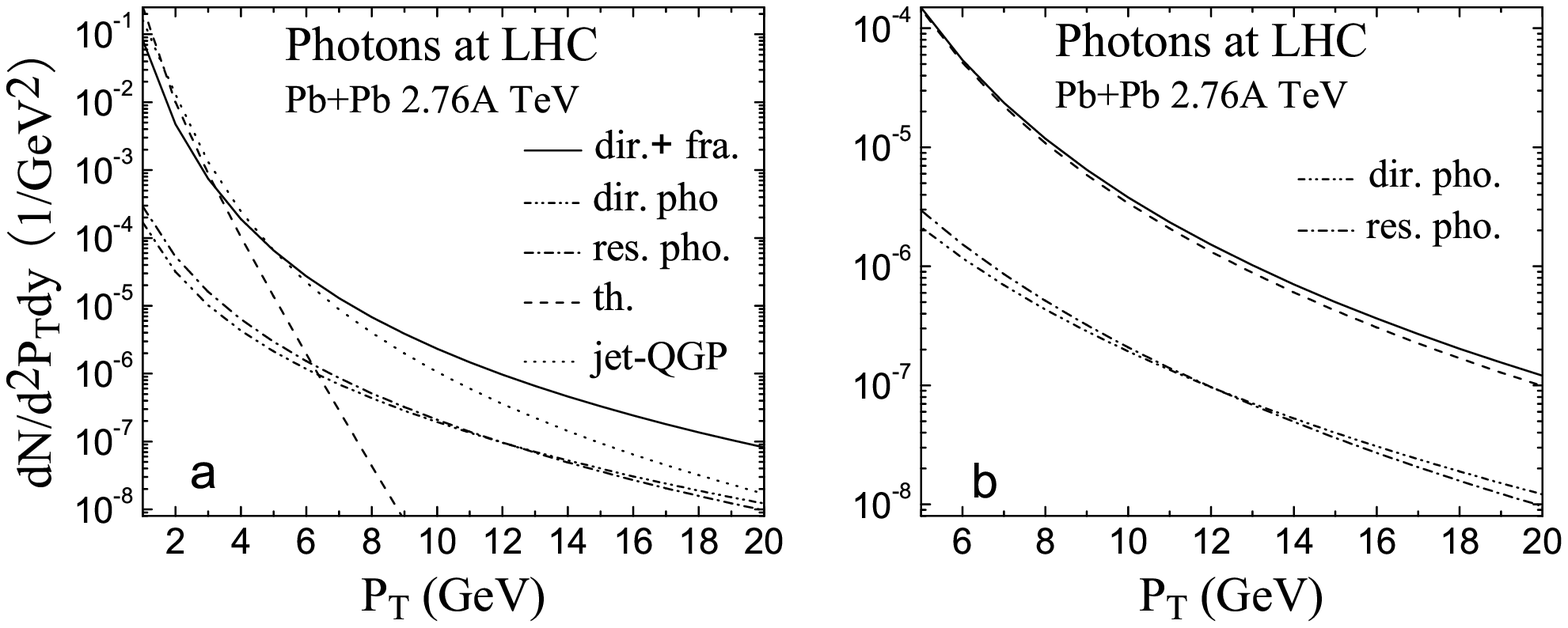}
\caption{\label{fig9}  Same as Fig.\ref{fig4} but for the real
photon production in central Pb+Pb collisions at
$\sqrt{s_{NN}}$=2.76 TeV.}
\end{figure*}


The rate of the photon production by Compton scattering and
annihilation of jets in the hot medium can be written as
\cite{18,19}
\begin{eqnarray}
E\frac{dR_{jet-\gamma}}{d^{3}P}\label{eq013}=\frac{\alpha\alpha_{s}e_{q}^{2}}{4\pi^{2}}
f_{jet}(\emph{\textbf{p}}_{\gamma})T^{2}\!\!
\left[2\ln\left(\frac{6E_{\gamma}}{\pi\alpha_{s}T}\right)+C\right],
\end{eqnarray}
where the constant is $C=C_{Com.}+C_{ann.}$. The detailed processes
of the thermal contribution and jet-$\gamma$($\gamma^{*}$)
conversion are discussed in Ref.\cite{19.1,19.2}. We briefly review
the contribution of thermal photons and dileptons. The
Landau-Pomeranchuk-Migdal (LPM) effect for the thermal production
and jet-$\gamma$($\gamma^{*}$) conversion \cite{19.1,19.2} is not
considered in our paper. In the calculation we use the Bjorken 1+1 D
evolution, the authors of Ref.\cite{19.1,19.2} consider the
transverse expansion of the hot and dense matter (3+1 D) in the
thermal photon and dilepton production.

\section{Numerical results}

The yield of large $P_{T}$ dileptons in a mass range between
$M_{min}$ and $M_{max}$ can be defined as \cite{3,4}
\begin{eqnarray}
\frac{dN_{AB\rightarrow
l^{+}l^{-}X}}{d^{2}P_{T}dy}=\int_{M_{min}}^{M_{max}}\frac{2M}{\pi}\frac{dN_{AB\rightarrow
l^{+}l^{-}X}}{dM^{2}dP^{2}_{T}dy}dM,
\end{eqnarray}
in this paper we choose the range 100 MeV$\leqslant M\leqslant $300
MeV.

\begin{figure*}
\includegraphics[scale=0.6]{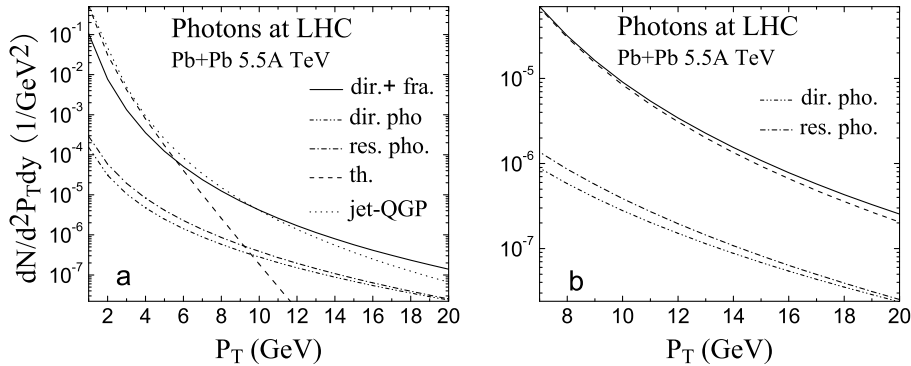}
\caption{\label{fig10}  Same as Fig.\ref{fig5} but for the real
photon production in central Pb+Pb collisions at $\sqrt{s_{NN}}$=5.5
TeV.}
\end{figure*}

The results of our calculation for $AA$ collisions in the minimum
bias case are plotted. In Fig.\ref{fig1} and \ref{fig2} we plot the
contribution of dileptons produced by direct and resolved
photoproduction processes for $pp$ and $dA$ collisions at RHIC and
LHC energies. In the panel a of Fig.\ref{fig1} and \ref{fig2} the
dilepton spectra of direct and resolved photoproduction
processes(dash dot line and dash dot dot line) are compared with the
spectrum of direct and fragmentation dileptons(dash line) for $p+p$
collisions and $d$+Au collisions at RHIC, respectively. We find that
the contribution of photoproduction processes is not prominent for
$p+p$ and $d$+Au collisions at RHIC energies. However,
photoproduction processes start playing an interesting role for
$p+p$ collisions and $d$+Pb collisions at LHC. The contribution of
photoproduction processes is evident in the region of $P_{T}>$1 GeV
for $p+p$ collisions(the panel b of Fig.\ref{fig1}), and $P_{T}>$1
GeV for $d$+Pb collisions at LHC energies(the panel b of
Fig.\ref{fig2}).

In the panel a of Fig.\ref{fig3} we plot the results for direct
dileptons(dir.), fragmentation dileptons(fra.), the jet-dilepton
conversion(jet-QGP) in the thermal plasma, and thermal
dileptons(th.) produced by the QGP in Au+Au collisions at RHIC. The
dilepton spectra of direct(dir. pho.) and resolved photoproduction
processes(res. pho.) are also plotted. The results for Pb+Pb
collisions at $\sqrt{s_{NN}}$=2.76 TeV and 5.5 TeV are shown in the
panel a of Fig.\ref{fig4} and the panel a of Fig.\ref{fig5},
respectively. In the panel b of Fig.\ref{fig3} we see that the
contribution of photoproduction processes is still weak for Au+Au
collisions at RHIC. However, the contribution of photoproduction
processes becomes evident in the large $P_{T}$ region at LHC
energies. In the panel b of Fig.\ref{fig4} and \ref{fig5} the
spectra of dileptons produced by direct and resolved photoproduction
processes(dash dot line and dash dot dot line) are compared with the
spectrum of direct dileptons, fragmentation dileptons, thermal
dileptons and the jet-dilepton conversion(dash line), we find that
the contribution of dileptons produced by photoproduction processes
is evident in the region of $P_{T}>$2 GeV for Pb+Pb collisions at
$\sqrt{s_{NN}}=$2.76 TeV, and $P_{T}>$4 GeV for Pb+Pb collisions at
$\sqrt{s_{NN}}=$5.5 TeV.

The contribution of real photons produced by direct and resolved
photoproduction processes is also negligible for $pp$, $dA$ and $AA$
collisions at RHIC energies(the panel a of Fig.\ref{fig6} and
\ref{fig7}; Fig.\ref{fig8}). However, the contribution of
photoproduction processes is evident in the region of $P_{T}>4$ GeV
for $p+p$ collisions at $\sqrt{s}$=7 TeV(the panel b of
Fig.\ref{fig6}), $P_{T}>4$ GeV for $d$+Pb collisions at
$\sqrt{s_{NN}}$=6.2 TeV(the panel b of Fig.\ref{fig7}), $P_{T}>5$
GeV for Pb+Pb collisions at $\sqrt{s_{NN}}$=2.76 TeV(the panel b of
Fig.\ref{fig9}), and $P_{T}>7$ GeV for Pb+Pb collisions at
$\sqrt{s_{NN}}$=5.5 TeV(the panel b of Fig.\ref{fig10}).

The photon spectrum $f_{\gamma/q}$ from the charged parton depends
on the collision energy $\sqrt{s_{NN}}$. We express the photon
spectrum as $f_{\gamma/q}\propto \ln\left((\hat{s}/4-m_{l}^{2})/ 1
GeV^{2}\right)=\ln(s_{NN}/1GeV^{2})+\ln(x_{a}x_{b}z_{a}.../4-m_{l}^{2}/s_{NN})$,
where $\hat{s}=x_{a}x_{b}z_{a}s_{NN}$ for direct photoproduction
processes and $\hat{s}=x_{a}x_{b}z_{a}z_{a'}s_{NN}$ for resolved
photoproduction processes. Since the collision energy at LHC is
larger than the collision energy at RHIC($s^{LHC}_{NN}\gg
s^{RHIC}_{NN}$), the photon spectrum becomes important at LHC
energies. Therefore the contribution of photoproduction processes is
evident at LHC.

We also plot the spectra of thermal dileptons and photons, because
the contribution of the thermal information is dominant in the small
$P_{T}$ region. We show the results of the jet-dilepton(photon)
conversion taking into account an effective 20$\%$ energy loss of
jets before conversion into dileptons(photons) \cite{12,19}. The
spectra of the jet-dilepton conversion fall off with the transverse
momentum of dileptons faster than the spectrum of primary hard
dileptons due to the attenuation function
$\exp(-P_{T}^{2}/4P'^{jet}_{T}T)$ in Eq.(\ref{eqjet01})(see the
panel a of Fig.\ref{fig3}, \ref{fig4} and \ref{fig5}). Since the
rate of the jet-photon conversion is $R_{jet-\gamma}\propto
f_{jet}$, the spectra of the jet-photon conversion do not drop
quickly with the transverse momentum(see the panel a of
Fig.\ref{fig8}, \ref{fig9} and \ref{fig10}).

\section{Summary}
We investigate the production of large $P_{T}$ dileptons and photons
in relativistic $pp$, $dA$ and $AA$ collisions by direct and
resolved photoproduction processes. In the initial parton scattering
the charged parton of the incident nucleon can emit large $P_{T}$
photons, then the high energy photons interact with the partons of
another incident nucleon by the QED Compton scattering. Furthermore,
the hadron-like photons also can interact with the partons of the
nucleon by the annihilation and Compton scattering. The numerical
results indicate that the contribution of photoproduction processes
is negligible for $pp$, $dA$ and $AA$ collisions at RHIC energies,
but the contribution becomes evident at LHC energies.

\section{Acknowledgements}
This work is supported by the National Natural Science Foundation of
China under Grant Nos 10665003 and 11065010.




\bibliographystyle{model1-num-names}
\bibliography{<your-bib-database>}


\end{document}